%% file: ms.tex
\documentclass{emulateapj}
\newcommand     \etal    {et al.}
\newcommand     \ergss  {\ensuremath{ergs~ s^{-1}}}
\newcommand	\lstar  {\ensuremath{L_{IR}^{*}}}
\shortauthors{Bai \etal}
\shorttitle{The IR Luminosity Functions of Rich Clusters}
\begin{document}
\title{The IR Luminosity Functions of Rich Clusters}
\author{Lei\,Bai\altaffilmark{1},
George H.~Rieke\altaffilmark{1},
Marcia J.~Rieke\altaffilmark{1},
Daniel Christlein\altaffilmark{2},
Ann I.~ Zabludoff\altaffilmark{1}}
\email{leibai@ociw.edu}
\altaffiltext{1}{Steward Observatory, University of Arizona, 933 N. Cherry Avenue, Tucson, AZ 85721}
\altaffiltext{2}{Max Planck Institute for Astrophysics, Garching, Germany}
\begin{abstract}
We present MIPS observations of the cluster A3266. 
About 100 spectroscopic cluster members have been detected at 24 \micron.
The IR luminosity function (LF) in A3266 is very similar to that in the Coma cluster down to the detection limit $L_{IR}\sim10^{43} \ergss$, suggesting a universal form of the bright end IR LF for local rich clusters with $M\sim10^{15}M_{\sun}$. 
The shape of the bright end of the A3266-Coma composite IR LF is not significantly different from that of nearby field galaxies, but the fraction of IR-bright galaxies (SFR $> 0.2M_{\sun}$ yr$^{-1}$) in both clusters increases with cluster-centric radius. 
The decrease of the blue galaxy fraction toward the high density cores only accounts for part of the trend; the fraction of red galaxies with moderate SFRs ($0.2~M_{\sun}$ yr$^{-1} <$ SFR $< 1~M_{\sun}$ yr$^{-1}$) also decreases with increasing galaxy density.
These results suggest that for the IR bright galaxies, nearby rich clusters are distinguished from the field by a lower star-forming galaxy fraction, but not by a change in $L_{IR}^{*}$.
The composite IR LF of Coma and A3266 shows strong evolution when compared with the composite IR LF of two $z\sim0.8$ clusters, MS 1054 and RX J0152, with $L^{*}_{IR}\propto (1+z)^{3.2^{+0.7}_{-0.7}},\Phi^{*}_{IR}\propto (1+z)^{1.7^{+1.0}_{-1.0}}$ .
This $L^{*}_{IR}$ evolution is indistinguishable from that in the field, and the $\Phi^{*}_{IR}$ evolution is stronger, but still consistent with that in the field. 
The similarity of the evolution of bright-end IR LF in very different cluster and field environments suggests either this evolution is driven by the mechanism that works in both environments, or clusters continually replenish their star-forming galaxies from the field, yielding an evolution in the IR LF that is similar to the field.
The mass-normalized integrated star formation rates (SFRs) of clusters within 0.5$R_{200}$ also evolve strongly with redshift, as $(1+z)^{5.3}$.
\end{abstract}


\section{INTRODUCTION}
Galaxy clusters are unique laboratories to study the environmental effects on galaxy evolution.
A high density environment could potentially alter the star formation (SF) in galaxies through many different physical mechanisms, e.g., ram pressure stripping, galaxy-galaxy interactions, and strangulation of the gas reservoir of galaxies \citep[see][and references therein]{Boselli06}.
As an extreme case of high density environment, clusters show clear evidence of less star formation compared to the field at both local and higher redshifts \citep{Kennicutt83,Balogh97,Hashimoto98,Poggianti99}, suggesting environmental mechanisms play a role in transforming their star forming properties \citep{Christlein05}.
However, it is still not clear which mechanism, or mechanisms, operate and where the transformation happens in the first place.
Is it an end product of galaxy `preprocessing' in the group environment \citep{Zabludoff96,Zabludoff98,Lewis02,Gomez03} or is it caused by some mechanism specifically related to the cluster environment \citep[c.f., ][]{Berrier08}?
Quantitative and unbiased star formation measurements in clusters and their comparison to field galaxies are keys to answering these questions. 
  
The luminosity function (LF) provides such a quantitative tool to study galaxy properties.
It has been widely used to compare galaxies in clusters with those in the field at many different wavelengths.
However, some of the results are controversial.
For example, some studies claim that cluster LFs show little variation across a wide range of cluster properties \citep{Colless89,Rauzy98,DePropris03,Popesso06}, while others find them to depend on cluster richness, Bautz-Morgan type \citep{Bautz70}, or distance from the cluster center \citep{Dressler78,Garilli99,Lopez97,Hansen05,Barkhouse07}.
Another question is whether cluster LFs differ from those in the field.
De Propris et al. (1998), \citet{Christlein03}, \citet{Cortese03} and \citet{Bai06} found cluster LFs to be indistinguishable from field LFs, but others suggest that they have brighter characteristic magnitudes and different faint end slopes \citep{Valotto97,Goto02,Yagi02,DePropris03}.

One of the major reasons for these apparent contradictions is that these studies are based on different wavelength ranges and therefore probe different properties of the cluster galaxy population.
The optical and near-IR luminosities of galaxies measure the old stellar mass, while the blue and UV band luminosities are more sensitive to the young stellar population. 
Since the LFs of different types of galaxies are different \citep{Madgwick02}, the LFs based on different wavelengths are biased toward different sub-populations of cluster galaxies.
\citet{DePropris03} found that although the B-band LF of early-type galaxies in clusters is different from that of the field, the B-band LF of late-type galaxies in clusters is very similar to that of the field.
A similar result was also found in the R-band for star forming and quiescent galaxies by \citet{Christlein03}.
\citet{Cortese05} found that the UV LF of star-forming galaxies in a nearby cluster is consistent with the field and that the difference between the cluster and field LFs is due to the quiescent galaxies. 

Although UV and B-band luminosities are sensitive to star-forming galaxies, they can be heavily attenuated by dust and therefore bias the LF.  
In contrast, the total IR luminosity ($L_{IR}$, 8-1000 \micron) shows a good correlation with the star formation rate (SFR) of galaxies unaffected by extinction \citep{Kennicutt98}. 
\citet{Bai06} studied the IR LF of the Coma cluster and concluded that its shape shows no significant difference from that of the field IR LF. 
However, this result is drawn from the study of one rich cluster and its validity in other clusters needs to be tested.

The study of the evolution of LFs with redshift can provide crucial constraints on models for the formation and evolution of galaxies.
In particular, the evolution of the IR LFs will help us understand the star formation history.  
By comparing Coma with two high redshift clusters, MS~1054-03 and RX~J0152 \citep{Marcillac07} , \citet{Bai07} found a strong evolution of the IR LFs for luminous galaxies in both density and luminosity (\lstar) from $z\sim0$ to $\sim0.8$. 
The evolution rate is consistent with that found in field IR LFs and suggests that this evolution pattern is probably not caused by cluster environmental mechanisms, but by processes similar to those that regulate the field IR LF evolution. 
However, because the comparison was based on only one low redshift cluster, Coma, there are possible systematic uncertainties in this evolution trend, especially if the IR LFs of local rich clusters have a large cluster-to-cluster variation. 
It has become clear that the Coma Cluster, once thought to be the archetype of a well-relaxed cluster, has many small substructures \citep{Mellier88,White93, Neumann03}.
Although we did not find any significant difference between the IR LF of the infalling NGC 4839 group and that elsewhere in the Coma cluster, it is still not clear how these substructures may affect the overall IR LF.
To study how different dynamical states may affect the IR LF and to quantify cluster-to-cluster variations, we need to study more low redshift clusters.

To test the universality of the IR LF in local clusters and provide a reliable baseline for both the cluster-field comparison and for evolution in clusters, we present the IR LF of A3266 in this paper.
A3266 is a nearby rich cluster ($z=0.06$) with a mass similar to Coma ($\sim10^{15}M_{\sun}$).
It has a large sample of spectroscopically confirmed cluster members (more than 300), making it possible to obtain a IR LF complete down to $L_{IR}\sim10^{43}$ \ergss. 
It is very bright in the X-ray ($L_{X}\approx10^{45}$ \ergss).
Its X-ray morphology and temperature map, as well as a comparison of the core and overall velocity dispersions, all suggest a recent major merger \citep{Quintana96,Henriksen00,Sauvageot05,Finoguenov06}. 
In \S 2, we present the data analysis for this cluster. 
In \S 3, we calculate the IR luminosities and SFRs for the cluster members.
In \S 4, we determine the IR LF for A3266, compare it with Coma's, and look at the dependence of the IR galaxy properties with cluster-centric radius.
In \S 5, we discuss the implications of the results and study the evolution of the integrated SFRs.  
In \S 6, we summarize the main conclusions.
Throughout this paper, we assume a $\Lambda$CDM cosmology with parameter set $(h,\Omega_{0},\Lambda_{0}) = (0.7,0.3,0.7)$.

\section{OBSERVATION AND DATA REDUCTION}
\subsection{MIPS Observations and Source Extraction}
The A3266 cluster was observed by MIPS \citep{Rieke04} on $Spitzer$ in medium scan map mode (GTO program \# 83, PI G. Rieke) on Jun 28, 2005 at 24, 70, 160 \micron\ simultaneously.
A rectangular $45\arcmin \times 60\arcmin$ region was mapped.
The data were processed with the MIPS Data Analysis Tool \citep[DAT version 3.02;][]{Gordon05} and array-averaged background subtraction was applied to improve the images.
The final mosaics have exposure times of $\sim80$, $\sim40$ and $\sim8$ s pixel$^{-1}$ at 24, 70 and 160 \micron, respectively.
The cluster resides in a region with low IR background ($\sim15 MJy/sr$).
The online SSC tool SENS\_PET estimates that the $3~\sigma$ sensitivity of these observations are 0.25, 15 and 150 mJy at the 24, 70 and 160 \micron\ bands respectively. 

We used SExtractor to extract sources and measure their photometric values.
Before source detection, the images were filtered with Gaussian functions with the full width at half maximum (FWHM) matching the FWHM of the point spread function of the images to help improve the detection of faint sources.
The flux of each source (flux\_auto) was computed in an adaptive Kron aperture.

\subsection{Spectroscopic Data}
To derive a spectroscopically-confirmed cluster member list, we retrieved galaxies with $0.047<z<0.072$ in the cluster field from the NASA/IPAC Extragalactic Database (NED).
This redshift range corresponds to $\pm3$ times the line-of-sight velocity dispersion of the cluster \citep{Christlein03}.
We retrieved 298 cluster members in the central $45\arcmin \times 60\arcmin$ region observed by MIPS. 

\begin{figure}
\epsscale{0.8}
\plotone{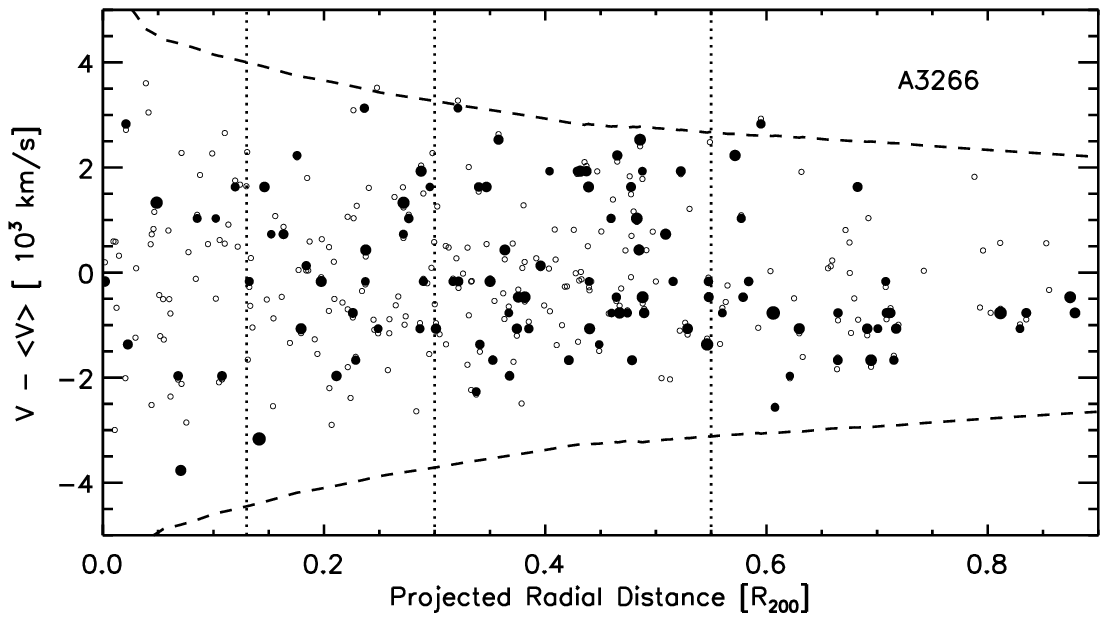}
\caption{The line-of-sight velocity relative to the cluster mean versus projected distance from cluster center.
The open circles are cluster members selected by $\pm3\times$ velocity dispersion, and the filled ones are those detected at 24 \micron.
The symbol sizes are proportional to the 24 \micron\ flux.
The two dashed curves are the most stringent criterion of the maximum line-of-sight velocity proposed by \citet{denHartog96}. All the members fall inside (or very close) to these curves.   
The three vertical lines at 0.13, 0.3 and 0.55 $R_{200}$ are the boundaries of the three regions used in \S 4.3 and \S 4.4.} 
\label{f_phaseplot}
\end{figure}

The simple method we used here for member selection is consistent with more sophisticated methods that combine radial velocity and position information.
In Fig.~\ref{f_phaseplot}, we plot the relative line-of-sight velocity ($V-\langle V \rangle$, $\langle V \rangle$ is the cluster mean) versus projected distance from the cluster center for all the members we select.
We also plot the maximum line-of-sight velocity as a function of the projected distance as proposed by \citet{denHartog96}.
All the members fall within, or very close to, this stringent criterion of member selection. 
We conclude that the simple velocity selection in the cluster region studied here works efficiently in excluding foreground/background galaxies.  

The spectroscopic data in NED come from several different surveys and are not homogeneous.
Therefore, we made use of a $R$-band photometric catalog in the cluster field \citep{Christlein03} to define the completeness of these data.
We obtained all the galaxies with redshifts in the same region from NED and compared their number with the total number of galaxies in the photometric catalog.
In Fig.~\ref{f_speccp}, we show these ratios as a function of $R$ magnitude.
As can be seen in the plot, the spectroscopic data have a high degree of completeness at $R<17$ mag, but the completeness drops rapidly to less than 50\% at $R>17.5$ mag.

This completeness function does not change much across the whole survey region, although the data are slightly more complete in the cluster center.
When we calculate the IR LF of the cluster members in the following section, we will use the inverse of this function as weighting factors to correct for the incompleteness of the spectroscopic survey.

\section{ANALYSIS}

\subsection{Correlating the IR emission with Cluster Members}
We correlated the spectroscopically confirmed cluster members with the nearest 24, 70 and 160 \micron\ sources projected within $6\arcsec$, $10\arcsec$ and $15\arcsec$, respectively.
These matching radii take into consideration the large FWHM of the images, $6\arcsec$, $18\arcsec$ and $40\arcsec$ at 24, 70 and 160 \micron, respectively.
At 24 \micron, the rather large matching radius relative to the small astrometric uncertainties ($<1\arcsec$) also accounts for the possible physical displacement between the optical and 24 \micron\ brightness centroids.
None of the 24 \micron\ sources above the $3\sigma$ detection limit has multiple matches with the spectroscopic cluster members.
Even when matching 24 \micron\ sources with all the extended sources in the photometric catalog, only 3\% of the mid-IR sources have more than one optical counterpart.
This demonstrates that blending is not an issue in correlating the IR emission with these cluster members.
Altogether, we found 109, 48 and 15 cluster members with 24, 70 and 160 \micron\ emission, respectively, above the $3\sigma$ detection limits.

\subsection{Deducing Total IR Luminosity from 24 \micron\ Emission}
The total IR luminosity ($\lambda=8-1000\micron$) of a star-forming galaxy is proportional to its SFR and can be used as a SFR indicator \citep[e.g.,][]{Kennicutt98}.
However, because direct measurement of the total IR luminosity is only possible for a limited number of galaxies, many SFR studies base their calculation on the total IR luminosity extrapolated from a single band IR measurement.
The mid-IR continuum emission, e.g., in {\it the Infrared Astronomical Satellite (IRAS)} 12 \micron\ and $Spitzer$ 24 \micron\ bands, has been shown to correlate with the total IR luminosity very well \citep{Takeuchi05,Alonso06,Calzetti07}.

For the sake of continuity with our previous work, we base our estimates of the total IR luminosities on a sample of star-forming galaxy SEDs developed by \citet{Dale02}.
These SEDs are based on $IRAS$ and {\it Infrared Space Observatory (ISO)} observations of 69 normal star-forming galaxies and have been well calibrated in the far-infrared and submillimeter bands.
They are luminosity-dependent and their total IR luminosities are monotonically correlated with the 24 \micron\ luminosity.
We find the IR luminosities of the cluster members from their 24 \micron\ flux by interpolating the $L_{IR}-f_{24}$ relation given by the template SEDs (see Table~\ref{tbl1}).
The deduced total IR luminosities of the cluster members above the $3\sigma$ 24 \micron\ detection level range from $4\times 10^{42}$ to $3.6\times 10^{44}$ \ergss. 

To test if the total IR luminosities deduced from the single 24 \micron\ band are consistent with the emission at longer wavelengths, we compare the 70 $vs.$ 24 \micron\ flux ratio ($f_{70}/f_{24}$) of the SED templates to the flux ratio of cluster galaxies detected at both wavelengths.
Because of the large uncertainties in the 70 \micron\ flux measurement of faint sources, we only include galaxies detected at 70 \micron\ with a significance $>6\sigma$.
For 43 galaxies with secure 24 and 70 \micron\ flux measurements, the average $f_{70}/f_{24}$ is 13, only slightly larger than the $f_{70}/f_{24}\sim12$ given by the SEDs.
We also determine the total IR luminosities from the observed 70 \micron\ flux by interpolating the $L_{IR}-f_{70}$ relation given by the template SEDs. 
The total IR luminosities deduced in this way are consistent with those deduced from the 24 \micron\ flux, with a small dispersion of $\sim0.1$ dex. 
This consistency confirms the validity of using the 24 \micron\ flux alone for the total IR luminosity determination.
Therefore, the following studies are all based on the 24 \micron\ data only.

\subsection{Comparison of SFRs Deduced from Different Methods}
Once we have the total IR luminosities of the galaxies, we calculate SFRs using the standard formula $SFR [M_{\sun} yr^{-1}]=4.5\times10^{-44}L_{IR} [\ergss]$ from \citet{Kennicutt98}.
Recent studies \citep{Alonso06,Pablo06,Calzetti07,Rieke08} have shown that the 24 \micron\ emission correlates tightly with SFR and suggest using the 24 \micron\ emission directly as a SFR indicator to minimize the uncertainties associated with deriving the total IR luminosity.
\citet{Rieke08} find that the 24 \micron\ flux provides accurate SFR estimates over a large infrared luminosity range.
For normal star-forming galaxies with $10^{10} L_{\sun} < L_{IR} < 10^{11} L_{\sun}$, they provide a relation SFR$(M_{\sun}yr^{-1})=7.8\times10^{-10}[L_{24 \micron}/L_{\sun}]$ accurate within 0.2 dex. 
We calculate the SFRs with this formula and compare them with the SFRs that we derived using our SED fitting routine.
We find good consistency between these two SFR estimates, with a systematic offset of $\sim0.3$ dex.
The systematic offset arises primarily from the different stellar initial mass functions (IMFs) used in \citet{Kennicutt98} and \citet{Rieke08}. 
The offset is reduced to $\sim0.1$ dex once allowance is made for the difference introduced by the IMFs.
We will use the SFR obtained from the Dale \& Helou templates in this paper. 
This approach facilitates comparison with previous work since systematic errors in different SFR derivations are avoided.

For the luminosity ranges of interest in this work, the UV-emitting (unobscured) component of star formation is a small fraction of the infrared-emitting one \citep[e.g.,][]{Bai07}, so we have not tried to correct for it. 
There is no reason to expect such a correction to evolve strongly with redshift over the $z\leq0.8$ range relevant to this paper. 
The infrared measurements as used here provide a single, uniform, and accurate metric for the SFR over the entire redshift and luminosity range of interest. 

\section{RESULTS}
\subsection{IR Luminosity Function of A3266}
\begin{figure}
\epsscale{0.6}
\plotone{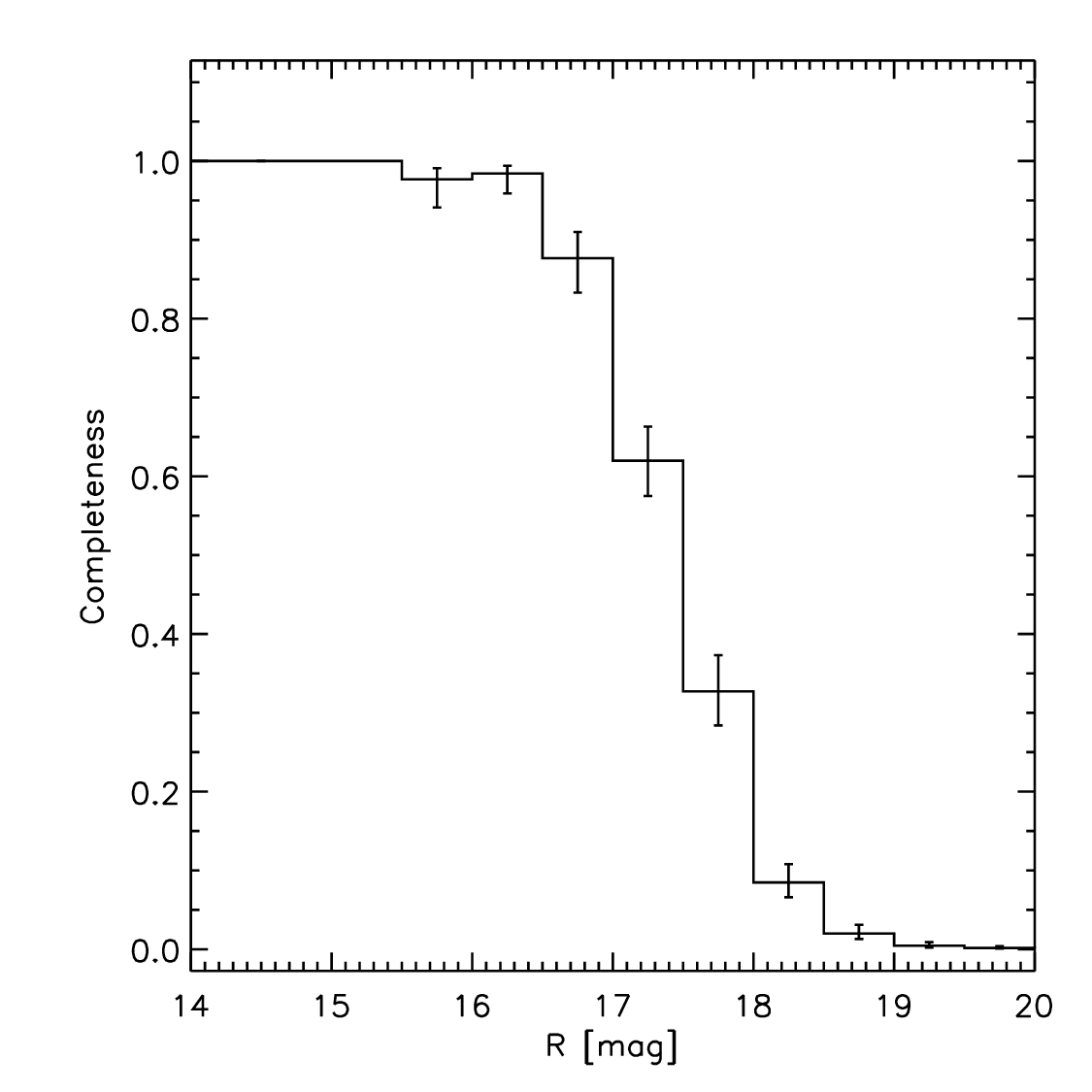}
\caption{ The completeness of the spectroscopic data in the cluster field as a function of $R$ magnitude.  The error bars are the $1\sigma$ errors given by binomial statistics.}
\label{f_speccp}
\end{figure}

\begin{figure}
\epsscale{0.6}
\plotone{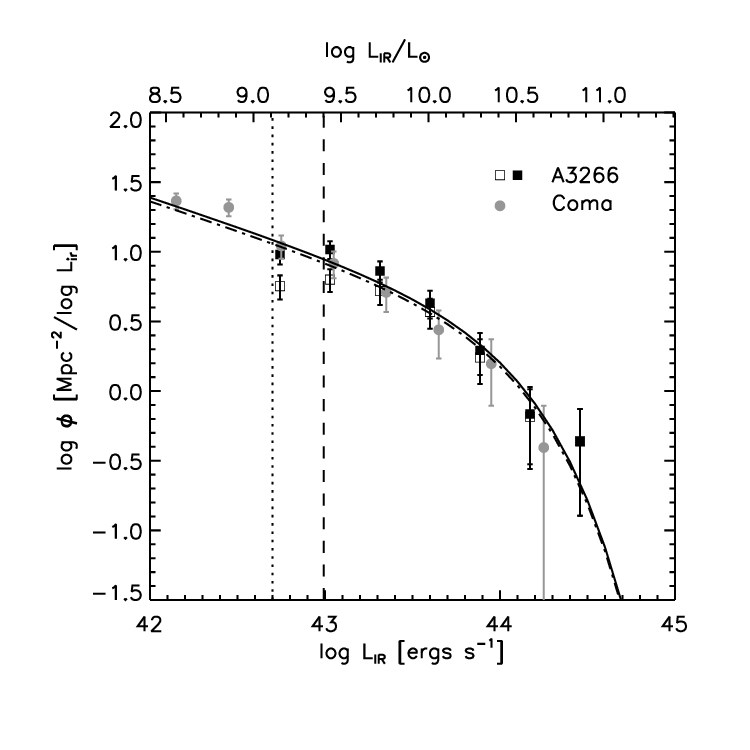}
\caption{The IF LF of A3266 (open squares).  The filled squares are the IR LF after correction for the incompleteness of the spectroscopic data and the solid curve is its best fitting Schechter function.  The gray filled circles are the IR LF of the Coma cluster \citep{Bai06} and the dash-dotted curve is its best fitting Schechter function.  The dotted vertical line is the luminosity corresponding to the detection limit of the 24 \micron\ data.  The dashed vertical line is the luminosity corresponding to the spectroscopic detection limit.}
\label{f_IRLF}
\end{figure}

We calculate the IR LF over the 16 Mpc$^2$ surveyed area in A3266.
The number of IR cluster members in each luminosity bin is divided by the surveyed area to obtain the surface density of IR cluster galaxies in that bin.  
The results are shown as the open squares in Fig.~\ref{f_IRLF}; the vertical dotted line is the 24 \micron\ detection limit.
To account for the incompleteness of the spectroscopic data, we use the inverse of the completeness function (shown in Fig.~\ref{f_speccp}) as weighting factors when we calculate the number counts in each luminosity bin.

The filled squares are the results of the incompleteness-corrected IR LF.
The correction raises the LF slightly at the faint end, but there is still a drop of galaxy number density at the lowest luminosity.
The drop is caused by the limits in the sensitivity of the spectroscopic survey. 
If we calculate the median $f_{R}/f_{24}$ ratio of the cluster members with $17 < R < 17.75$, where the spectroscopy is around 50\% complete, and assume that the fainter galaxies have similar $f_{R}/f_{24}$ ratios, the median $L_{IR}$ is about $10^{42.7}$ \ergss\ for galaxies at $R=19$.
Although the IR data are complete to this luminosity level, only about 5\% of the cluster members have spectroscopy (Fig.~\ref{f_speccp}).
Thus, we start to lose IR galaxy number counts due to the detection limit of the spectroscopic survey, explaining the drop in the lowest luminosity bin.
Another problem of the simple strategy we adopt to correct the incompleteness is the color bias between the cluster and field galaxies. 
We calculate the spectroscopic completeness in the $R$ band, but at least 70\% of the spectroscopic targets are selected in the $B$ band \citep{Christlein03}.
Because field galaxies are generally bluer than cluster galaxies, this color bias will cause an overestimate of the spectroscopic completeness in the $R$ band for cluster galaxies and therefore an underestimate of the number density of IR cluster galaxies.    
We expect that this color bias does not affect the data points brighter than $L_{IR}\approx10^{43}$ \ergss, because most of these IR bright galaxies are blue in color too.
However, many of the IR galaxies with $L_{IR}<10^{43}$ \ergss\ are in the red sequence (as suggested by the Coma cluster data), and the spectroscopic incompleteness correction that we make may not be adequate there.

Along with the IR LFs of A3266, we plot the IR LF of the Coma cluster \citep{Bai06} in Fig.~\ref{f_IRLF}.
The Coma IR LF is obtained with a similar data set as in A3266.
However, \citet{Bai06} use slightly different SEDs to deduce the total IR luminosity from the 24 \micron\ flux density.
We update the total IR luminosities of the Coma galaxies using the same SEDs that we use in this paper.
The resulting Coma IR LF is only slightly changed.

From the plot, we can see that the IR LF of A3266 is very similar to that of the Coma cluster in the luminosity region above the completeness limit.
For further comparison, we fit the incompleteness-corrected IR LF of A3266 with a Schechter function \citep[$\phi\propto(L/L^{*})^{-\alpha}e^{-L/L^{*}}$,][]{Schechter76}.
We adopt the chi-square minimization method used in \citet{Bai06,Bai07} to find the best fitted parameters.
The method incorporates the non-detection of galaxies beyond the brightest bin into the chi-square calculation. 
Because we do not have many data points to constrain the faint end slope, we have fixed it at $\alpha=1.41$, the value given by the Coma IR LF.
We exclude the faintest data point, which is affected by the inadequate incompleteness correction, from the fitting.
The resulting best fitted parameters are:
\begin{equation}
\alpha=1.41~(fixed),~{\rm log}(L^{*}_{IR}/L_{\sun})=10.49^{+0.13}_{-0.11}.\
\end{equation} 
The $L^{*}_{IR}$ value is the same as the value found in the Coma cluster \citep[${\rm log}(L^{*}_{IR}/L_{\sun})=10.49^{+0.27}_{-0.24}$,][]{Bai06}.
The close match in both LF shape and $L^{*}_{IR}$ demonstrates that the bright-end IR LFs are virtually identical. 
A similar result has been obtained previously for two high redshift clusters \citep[$z= 0.83$;][]{Bai07}, suggesting that the variation of the LF from one massive cluster to another at the same redshift is likely to be small. 

\subsection{Composite IR Luminosity Function\label{sect_cLF}}
\begin{figure}
\epsscale{0.6}
\plotone{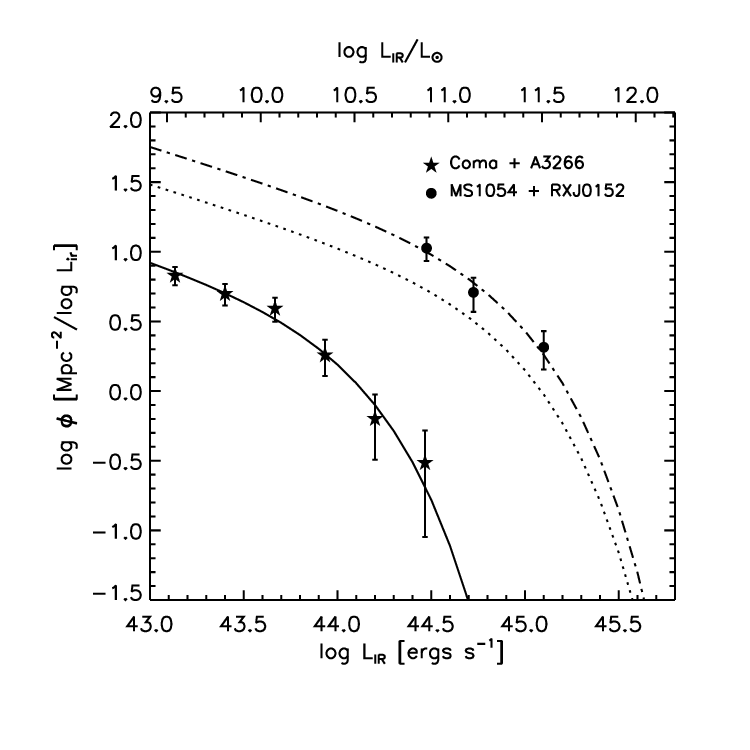}
\caption{The composite IR LF of A3266 and Coma (filled stars).  The best fitted Schechter function is shown as the solid curve. 
The filled circles are the composite IR LFs of MS~1054-03 ($z=0.83$) and RX~J0152 ($z=0.84$).  The dash-dotted line is its best fitted Schechter function.  The dotted line is the IR LF of A3266 and Coma evolved to $z=0.83$ using $L^{*}_{IR} \propto (1+z)^{3.2^{+0.7}_{-0.2}},~\Phi^{*}_{IR}\propto (1+z)^{0.7^{+0.2}_{-0.6}}$ \citep{LeFloch05}.}
\label{f_compositeLF}
\end{figure}

The similarity of the IR LFs of the Coma cluster and A3266 allows us to obtain a composite LF for these two clusters.
For this purpose, we include only the IR cluster members with $L_{IR} > 10^{42.7}$ \ergss$^{-1}$.
Above this limit, the IR surveys in these two clusters are both nearly complete, and the spectroscopic incompleteness correction in A3266 works well.
The composite IR LF is shown in Fig.~\ref{f_compositeLF}.
The best fitted parameters of the Schechter function  are:
\begin{equation}
\alpha=1.41~(fixed),~{\rm log}(L^{*}_{IR}/L_{\sun})=10.50^{+0.12}_{-0.11}.\
\end{equation}
The $L^{*}_{IR}$ of the composite LF is well within the 1$\sigma$ error of the values obtained for the Coma and A3266 clusters individually.

\citet{Bai07} found the IR LF of a high redshift cluster, MS~1054-03 ($z=0.83$), shows strong evolution compared with the IR LF of the Coma cluster.
The best fitting Schechter function of MS~1054-03's IR LF has a $L^{*}_{IR}$ about an order of magnitude larger than that found in the Coma cluster.
To further test this trend, here we combine the data from MS~1054-03 with the data from another high redshift cluster, RX~J0152 \citep{Marcillac07}, and obtain a composite IR LF.
The total IR luminosities are deduced in the same way to minimize the difference caused by systematic errors.
The composite IR LF of the high-$z$ clusters has log$\lstar/L_{\sun}=11.35^{+0.14}_{-0.14}$ with a fixed $\alpha= 1.41$, very close to the value obtained for the IR LF of MS~1054-03 alone. 
The normalization is slightly smaller than the IR LF of MS~1054-03, but is still much higher than the composite LF of the low-$z$ clusters.
The \lstar\ value increases by almost an order of magnitude from $z\sim0$ to 0.8.
This result confirms the strong evolution of IR LFs that we found by comparing only the Coma IR LF to that of MS~1054-03.
A strong evolution is also found in the field IR LF from $z=0$ up to $z\sim1$.
\citet{LeFloch05} characterize the evolution of IR LF in the CDF-S field as $L^{*}_{IR}\propto (1+z)^{3.2^{+0.7}_{-0.2}},\Phi^{*}_{IR}\propto (1+z)^{0.7^{+0.2}_{-0.6}}$ up to $z\sim1$; \citet{Pablo05} found a similar result using a larger data set.  
In comparison, the evolution found between the low and high-$z$ cluster LFs can be described by $L^{*}_{IR}\propto (1+z)^{3.2^{+0.7}_{-0.7}},\Phi^{*}_{IR}\propto (1+z)^{1.7^{+1.0}_{-1.0}}$.
The evolution of the $L^{*}_{IR}$ in clusters agrees with the evolution in the field very well. 
The evolution of the normalization ($\Phi^{*}_{IR}$) in clusters is stronger than that in the field, but it is consistent with the field given the error of the measurement. 
In Fig~\ref{f_compositeLF}, we compare the evolution of IR LFs in clusters directly to that in the field by evolving the composite IR LF of Coma and A3266 to $z=0.83$ using the evolution trend given by the field IR LF.
As shown in the plot, the evolved LF is similar in shape to that observed in the two high-$z$ clusters, but with a smaller normalization.
Using H$\alpha$ emission lines, \citet{Finn08} studied the star formation properties of a large sample of local clusters and compared it with several $z\sim0.75$ clusters.
Consistent with our results, they found that the total H$\alpha$ luminosity of the high-$z$ clusters is 10x that of the local clusters and this evolution is comparable to the  H$\alpha$ luminosity evolution found in the field.

It is worth noting that the difference of IR LFs between our high-$z$ and low-$z$ cluster samples is primarily due to evolution, not to the difference in other cluster properties, e.g., mass.
Although these four clusters are all among the richest clusters at their epochs, the two low-$z$ clusters are more massive than the two high-$z$ clusters.
However, the mass of the Coma is only slightly greater than that of MS~1054-03 ($1.4\times10^{15}$ M$_{\sun}$ vs. $1.1\times10^{15}$ M$_{\sun}$).
A3266 is twice as massive as Coma, but its IR LF is very similar to Coma's; RX~J0152 is half of the mass of MS~1054, but again, their IR LFs are not very different \citep{Bai07}.
The similarity suggests that at the same epoch, the masses of these rich clusters do not affect the properties of their IR LFs substantially.
Although some evidence of a correlation between star formation and cluster mass has been found \citep{Poggianti06}, star formation is almost independent of cluster mass for the mass range probed here \citep{Finn08}.

Another issue that could bias the comparison is the different aperture of the clusters within which we calculate their IR LFs.
Since the cluster IR LF is the surface density averaged over the survey area, the different data coverage in each cluster could cause differences in their LFs \citep{Popesso06}.
For the three rich clusters we study here, Coma, A3266 and MS~1054-03, the IR LFs are all calculated in a region about 70\% of the $R_{200}$ area ($R_{200}$ is the radius within which the mean cluster density is 200 times the critical density of the universe at that redshift).
The data on RX~J0152 covered the whole $R_{200}$ area, however, its IR LF does not differ much from that of MS~1054-03 \citep{Bai07}.
In addition, in the following section (\S \ref{sect_LFDR}), we show that the IR LFs over the majority of the volumes of the two local clusters are consistent with their overall IR LFs. 
This result indicates that the aperture difference does not affect the evolution we find between high-$z$ and low-$z$ cluster IR LFs.

\subsection{Composite IR LFs at Different Radii \label{sect_LFDR}}
\begin{figure}
\epsscale{1.0}
\plotone{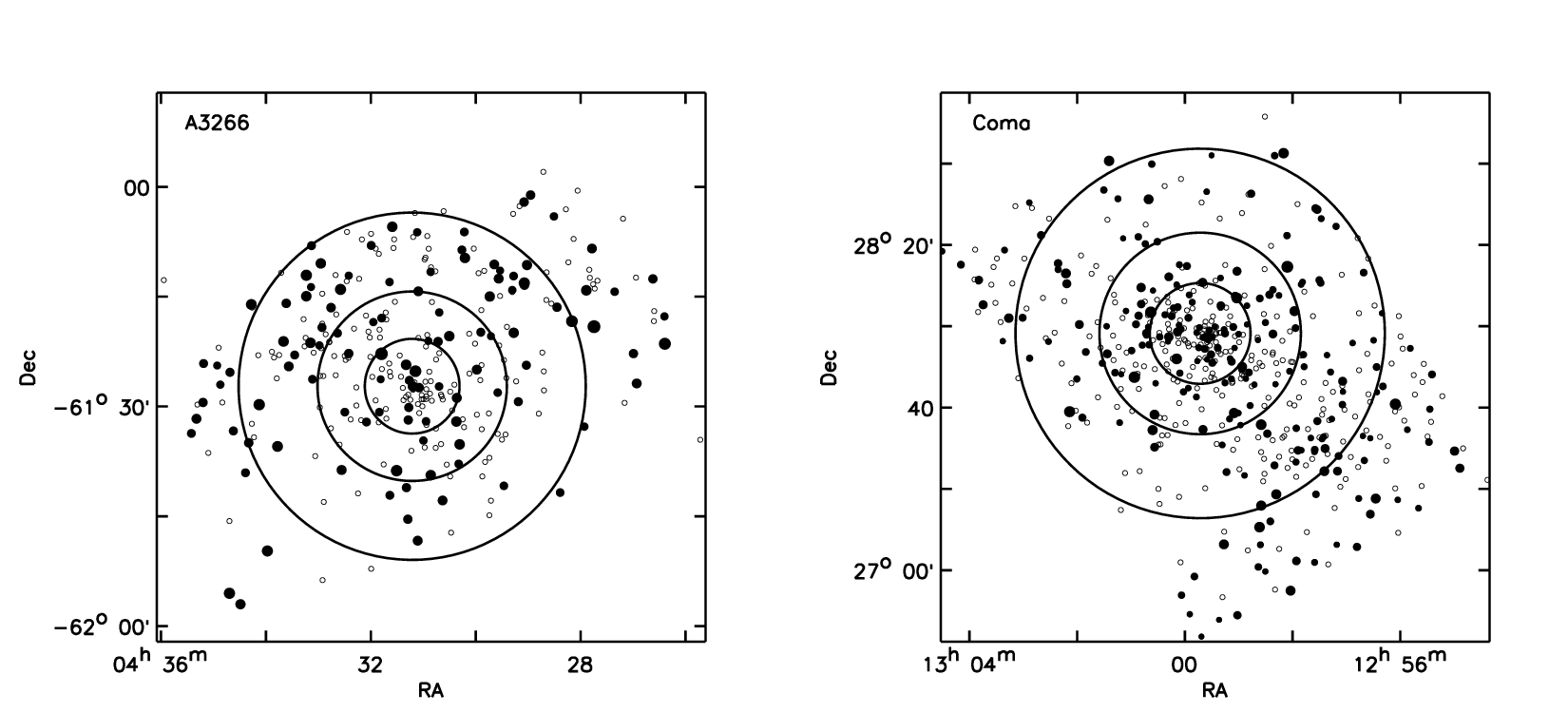}
\caption{The sky map of A3266 and Coma.  East is to the left and north is at the top.  The IR cluster members are shown as filled circles.  The symbol sizes are proportional to the 24 \micron\ flux.  The open circles are spectroscopically confirmed cluster members.  The three big circles indicate the three regions within which we extracted IR LFs ($r=0.13, 0.3,0.55R_{200}$).}
\label{f_IR_dist}
\end{figure}

\citet{Bai06} studied the IR LFs in different regions of the Coma cluster and found a flatter faint end slope and lack of very bright galaxies in the core region. 
Although the data in A3266 are not deep enough to constrain the faint end slope of its LF, the composite LFs combining similar radial regions of these two clusters have better statistics at the bright end and can constrain variations of the bright-end LF at different cluster radii.
For this purpose, we combine the same regions of the two clusters relative to $R_{200}$.
$R_{200}$ is better than a fixed physical scale because it takes into account the differences in the masses of the clusters \citep{Popesso06,Barkhouse07}.
We define three regions in each cluster: the core region ($r \leq 0.13R_{200}$), the intermediate region ($0.13R_{200} < r \leq 0.3R_{200}$), and the outer region ($0.3R_{200} <r \leq 0.55R_{200}$). 
The core region is chosen to match that defined earlier in the Coma cluster \citep[$\sim0.3$ Mpc,][]{Bai06}. 
The largest radius in the outer region is limited by the data coverage.
The three regions in A3266 and Coma are shown in Fig.~\ref{f_IR_dist}.
Altogether, there are 8, 23 and 44 IR galaxies with $L_{IR}> 10^{43}$ \ergss\ found in the core, intermediate, and outer regions, respectively.  

\begin{figure}
\epsscale{0.6}
\plotone{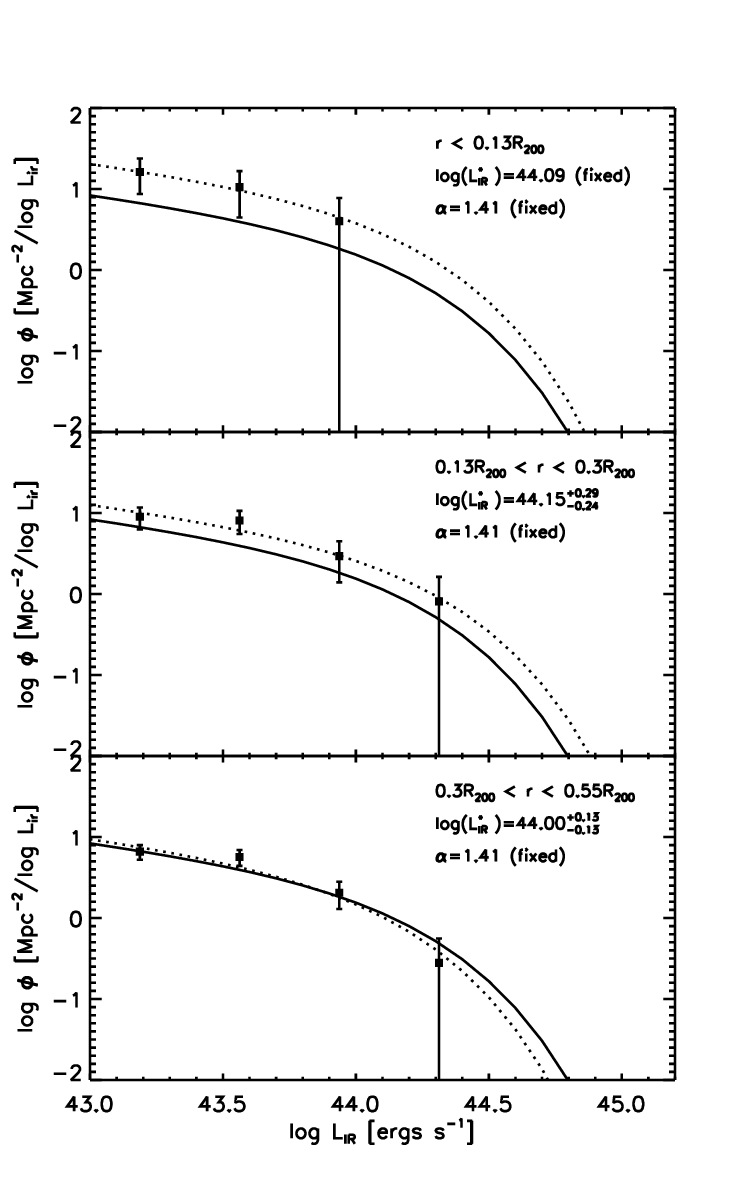}
\caption{The composite IR LFs in different regions of the clusters.  The solid curves are the best fitted Schechter function of the total composite IR LF of these two clusters.  The dotted curves are the best fitted Schechter function in the different regions.}
\label{f_LF_region}
\end{figure}

The resulting composite IR LFs in the three different regions are shown in Fig.~\ref{f_LF_region}.
In the core region, despite the high overdensity of cluster members, there are only 8 of them with $L_{IR}>10^{43}$ \ergss, and no galaxies in the brightest bin of the LF.
Because the data points are so few in this region, we can not derive a reasonable $L_{IR}^{*}$ value.
Therefore, we fix both $L_{IR}^{*}$ and $\alpha$ to the values of the total composite IR LF from \S \ref{sect_cLF} and only allow the normalization to change.
The best-fitted Schechter function has a slightly larger normalization than that of the total composite IR LF.
To further test whether those IR bright galaxies really reside in the cluster cores or are galaxies at larger radii being projected on to these regions, we compare their absolute relative line-of-sight velocity ($|V-\langle V\rangle|$) with the core galaxies that are not IR bright (see Fig.~\ref{f_phaseplot}).
In the A3266 core, the median value of $|V-\langle V\rangle|$ for the IR bright galaxies is $\sim1600$ km s$^{-1}$, much larger than the median of non-IR bright galaxies ($\sim 800$ km s$^{-1}$).
In Coma, there are only two galaxies with $L_{IR}>10^{43}$ \ergss\ in the core region, and the bright one has an absolute relative velocity about 2.5 times larger than the median value of the rest of the core members.
These results suggest that the few core galaxies with $L_{IR}>10^{43}$ \ergss\ are likely either to be projected on the core, or to be on orbits systematically different from those of the other core galaxies.
In either case, these galaxies are probably not among the well-relaxed core galaxy population. 

In the intermediate and outer regions, we fit the composite LFs with a Schechter function assuming a fixed faint end slope as before.
The difference between the intermediate region (log$L^{*}_{IR}=44.15^{+0.29}_{-0.24}$) and the outer region (log$L^{*}_{IR}=44.00^{+0.13}_{-0.13}$) is within the $1\sigma$ error, and both $L^{*}_{IR}$ values are consistent with the value for the total composite IR LF.
Limited by the small number counts, we are unable to perform a statistically significant comparison between the core IR LF and the LFs in other regions. 
Overall, except for the quite uncertain core IR LF, the IR LFs in different regions of the clusters are generally consistent with each other and with the overall composite IR LF.

\subsection{Fractions of IR-bright Galaxies}
Although the IR LFs of Coma and A3266 are almost identical at the bright end, the fractions of IR-bright galaxies in these two clusters are slightly different.
We calculate the fractions for all the galaxies with $M_{R}\le -20.15$.
This magnitude corresponds to $R=17.0$ and $14.9$ in A3266 and Coma, respectively, both above the 60\% spectroscopic completeness limits in their fields. 
In A3266, we find that $40^{+3}_{-4}\%$ of cluster members brighter than $M_{R}=-20.15$ have $L_{IR}>10^{42.7}$ \ergss\ (SFR$> 0.2~M_{\sun}~yr^{-1}$) and in Coma this fraction is $31^{+4}_{-5}\%$.
If we include the galaxies one magnitude fainter ($M_{R}\le -19.15$), the fractions are lower, $35^{+3}_{-2}\%$ and $23^{+3}_{-2}\%$ for A3266 and Coma, respectively.
This fainter magnitude cut includes all (except a couple) of the galaxies with $L_{IR}>10^{42.7}$ \ergss\ in both clusters and it corresponds to a galaxy stellar mass of $3-7\times10^{9}~M_{\sun}$ \citep{Bell01}. 
In A3266 this magnitude extends to the range where the spectroscopic data are rather incomplete and the calculation might not be reliable. 
We note that although A3266 has a higher fraction, and also a higher number, of IR-bright galaxies than Coma, they are distributed over a larger area and therefore the normalization of its IR LF is still similar to that in Coma.

\begin{figure}
\epsscale{0.6}
\plotone{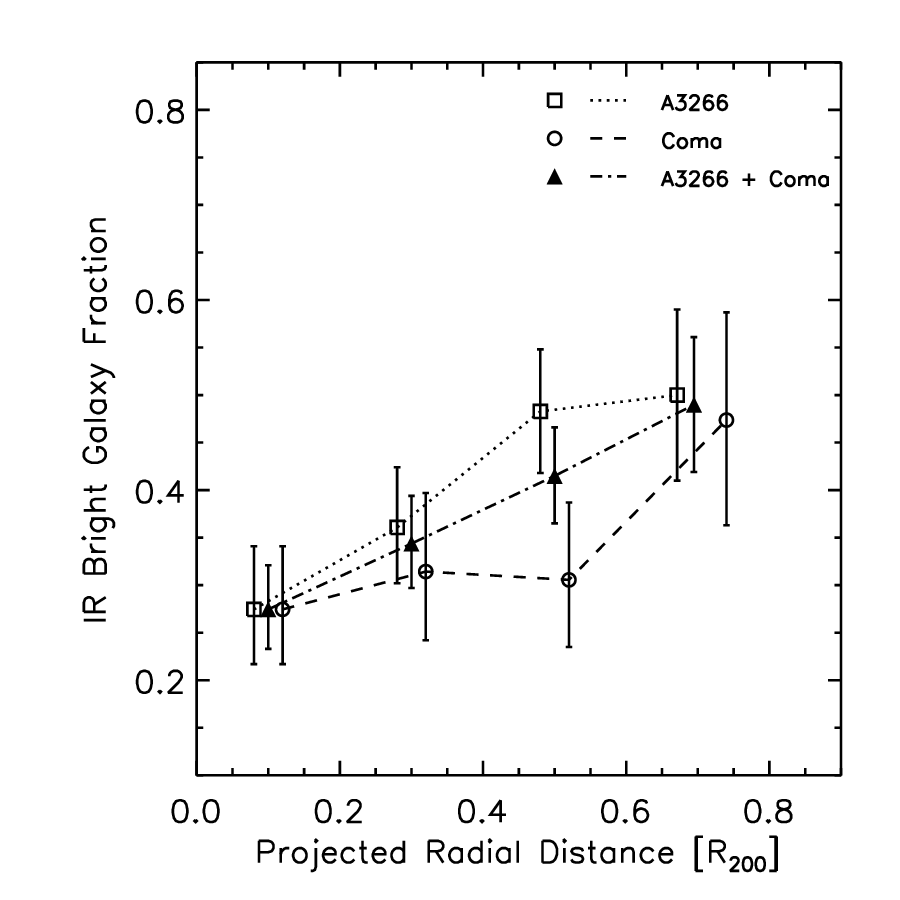}
\caption{The fraction of star-forming galaxies (log$L_{IR}>42.7$, $\sim SFR > 0.2~M_{\sun}~ yr^{-1}$) as a function of projected distance from the cluster centers. The fractions are only calculated for galaxies with $M_{R} \le -20.15$. The error bars are the $1\sigma$ errors given by binomial statistics. }
\label{f_fraction}
\end{figure}

From the velocity vs. projected radial distance plot (Fig.~\ref{f_phaseplot}) and the sky map of A3266 (Fig.~\ref{f_IR_dist}), we can see that these IR-bright galaxies scatter around the whole region, but appear to be more frequent at large radii, suggesting a change of IR-bright galaxy fraction with radius.
To quantify this trend, we plot the fraction of IR-bright galaxies at different projected radii in units of $R_{200}$.
For this fraction calculation, we use the brighter $R$ magnitude limit ($M_{R}\le -20.15$).
As shown in Fig.~\ref{f_fraction}, the fractions in A3266 are similar to those in Coma, except for the point at $R\sim0.5R_{200}$, where there are $\sim20\%$ more IR-bright galaxies in A3266 than in Coma.

Despite this difference, both clusters show a general increase of IR-bright galaxy fractions from the central to the outer region.
To improve the statistics of this general trend, we combine the data from the two clusters and find that the fractions increase smoothly with radius, with a linear relation, $(0.24\pm0.02) + (0.36\pm0.06)R/R_{200}$.
This general trend is consistent with the observed decrease of star-forming galaxy fractions with density \citep[e.g.,][]{Lewis02}.
In the Coma cluster, for which $B$ band photometry is available, we also calculate the blue galaxy fraction as a function of radius.
The blue galaxies are defined as galaxies with $B-R<1.95-0.05R$, 0.18 mag bluer than the mean color of the red sequence galaxies.
Among cluster members with $M_{R}<-20.15$, the limit used to calculate the IR-bright galaxy fraction, the blue galaxy fraction increases from $\sim6\%$ in the central region to $\sim15\%$ at $0.7R_{200}$.
These fractions are 4-5 times smaller than the IR-bright galaxy fractions.
However, if we increase the lower IR luminosity cut to $L_{IR}>10^{43.35}$ \ergss (SFR $>1M_{\sun}yr^{-1}$), the IR-bright galaxy fractions and the blue galaxy fractions match each other quite well. 
The colors of those IR-bright galaxies with SFR $>1M_{\sun}yr^{-1}$ confirm that the majority are blue galaxies. 
This indicates that the decrease of the blue galaxy fraction with galaxy density can only account for part of the decrease of the IR-bright galaxy fraction, namely, just for those galaxies with SFR $>1$ M$_{\sun}$ yr$^{-1}$, and that the fraction of the red galaxies with moderate SFR (up to $1M_{\sun}$ yr$^{-1}$) also decreases with galaxy density.

\subsection{The Evolution of the Integrated SFRs}
In \citet{Bai07}, we studied the integrated SFRs ($\Sigma SFRs$) of more than 10 clusters and found some tentative correlations with redshifts and masses. 
To test those results, here we add the data from A3266.
We integrate the SFRs of the cluster members within the 0.5$R_{200}$ ($\sim$ 1.5 Mpc) down to a SFR limit of 2 $M_{\sun}~ yr^{-1}$.
This gives a value of 39 $M_{\sun}~ yr^{-1}$.
The spectroscopic incompleteness correction slightly increases this value to 43 $M_{\sun}~yr^{-1}$.
The radius (0.5$R_{200}$) and the SFR limit (2 M$_{\sun}$ yr$^{-1}$) for calculating the $\Sigma SFRs$ are chosen to accommodate most of the cluster data in the sample.
In A3266, this integrated SFR only accounts for a quarter of the total integrated SFR of all the cluster galaxies that we study. 
We also calculate the mass-normalized $\Sigma$SFR ($\Sigma SFR/M$), for which we use a cluster mass of $3.3\pm{0.1}\times10^{15}~M_{\sun}$. 
The cluster mass is deduced using a velocity dispersion of 1255 km s$^{-1}$ \citep{Christlein03} and utilizing the formula given by \citet{Finn04}. 
The error on the mass takes into consideration the enhanced velocity dispersion caused by a possible merger \citep{Quintana96}.

\begin{figure}
\epsscale{1.0}
\plotone{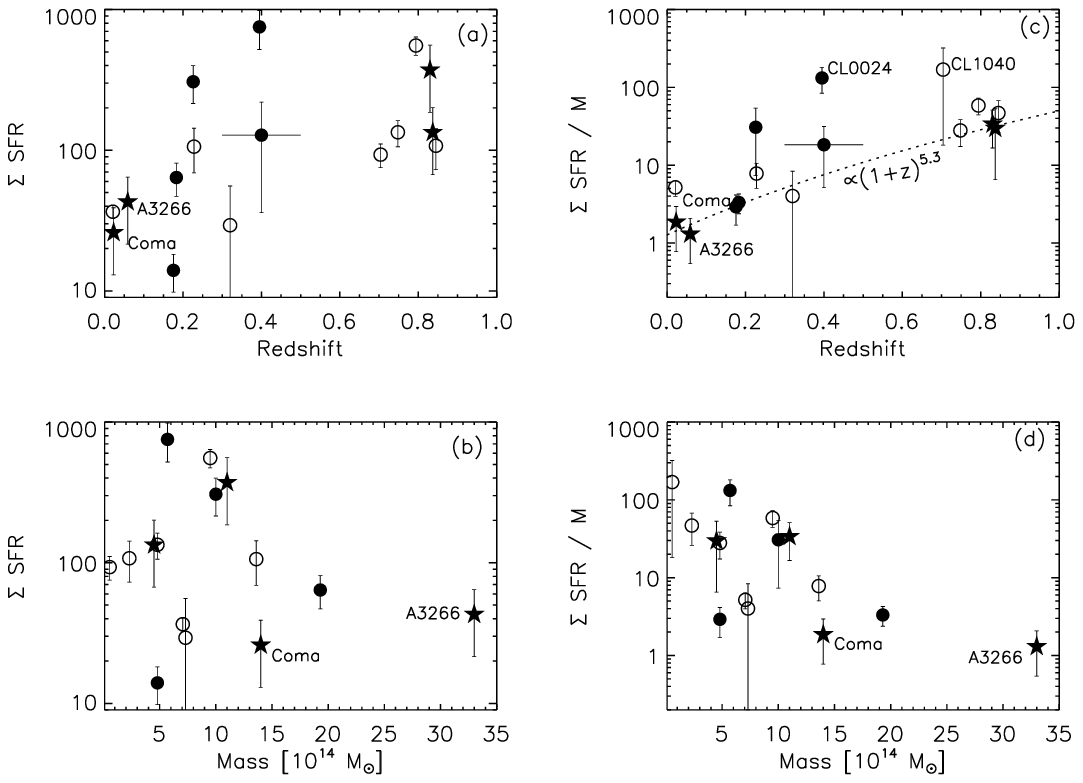}
\caption{($a$) and ($b$), the integrated SFRs ($<0.5R_{200}$) vs. redshifts and cluster masses;
($c$) and ($d$), the mass-normalized SFRs ($<0.5R_{200}$) vs. redshifts and cluster masses.
The filled stars are four clusters observed with MIPS: Coma, A3266, MS 1054-03, and RX~J0152.
Filled circles are the clusters observed with ISOCAM, and open circles are from H${\alpha}$ emission line measurements, adjusted as described in \citet{Bai07}.
The dotted curve in ($c$) is the fitted correlation between mass-normalized SFRs and redshifts for the four clusters observed by MIPS.
}
\label{f_A3266_intsfr}
\end{figure}

As can be seen in Fig.~\ref{f_A3266_intsfr}, both $\Sigma SFR$ and $\Sigma SFR/M$ in A3266 are consistent with the evolution trend that we found before. 
On the other hand, the relation between the $\Sigma SFR/M$ and cluster mass is less conclusive compared to the evolution trend.
Especially for the less massive clusters, there is large scatter in both the $\Sigma SFR$ and the $\Sigma SFR/M$.
The four rich clusters studied by MIPS, despite the large mass range (see Panel $d$ of Fig.~\ref{f_A3266_intsfr}), show good agreement in $\Sigma SFR/M$ at the same redshift. 
This argues against a strong dependence of $\Sigma SFR/M$ on cluster mass for these rich clusters.
Similarly, the IR LFs of these four rich clusters also argue against a strong dependence on cluster mass.
This is consistent with the recent results of \citet{Finn08}, who suggest that the difference between the fraction of star forming galaxies in the field and clusters first appears in clusters with velocity dispersions lower than 600 km s$^{-1}$ (ie, groups); clusters with higher velocity dispersions all have similar, and much lower than the field, fractions of star forming galaxies.

Using only the data points of these four clusters, for which we are most confident, we derive the evolution in the trend $\Sigma SFR/M \propto (1+z)^{5.3\pm1.2}$.
This relatively well-constrained evolution trend deduced from the richest clusters at two different epochs, lies under a more scattered evolution traced by other data points from ISO observations and $H_{\alpha}$ surveys.
The scatter may be partly due to the uncertainties of the aperture correction, detection limit correction and extinction correction (for $H_{\alpha}$ data) we made to obtain $\Sigma SFR$ \citep[see details in ][]{Bai07}, but more likely is due to some intrinsic difference in cluster properties that may affect their star formation properties.
For example, one of the two clusters deviating most from the evolution trend, CL~1040 ($z=0.7$) has a much smaller mass \citep[$\sim5\times 10^{13} M_{\sun}$, ][]{Clowe06} compared to other clusters in the sample and therefore probably has properties more similar to galaxy groups than clusters.
The other one, CL~0024 ($z=0.4$), is X-ray underluminous compared to the clusters with similar mass, suggesting a lack of hot intracluster medium (ICM) that may relate to the abnormally high star formation activity.
To disentangle all these effects from the evolution, we need not only a comparison of a well-defined and much larger cluster sample at different redshifts, but also comparison of cluster and group samples.

A recent study by \citet{Saintonge08} found that the fraction of star forming galaxies in eight massive clusters ($M>5\times10^{14}M_{\sun}$) increases from 3\% at $z=0.02$ to 13\% at $z=0.83$. 
The SFR limit cut ($>4$ M$_{\sun}$ yr$^{-1}$) and the region ($<$1 Mpc) they used to calculate the fractions are comparable to what we used to calculate the $\Sigma SFR$. 
The star-forming galaxy fraction evolution, $\sim(1+z)^{2.4}$, is not as strong as the $\Sigma SFR/M$ evolution we find here. 
This is because the evolution of $\Sigma SFR/M$ also takes into account the evolution of $L^{*}_{IR}$ for star-forming galaxies.
This again confirms what we find by studying the evolution of the IR LFs: both the density and the $L_{IR}$ of the star-forming galaxies have evolved strongly from $z=0$ to $z=0.8$ in rich clusters.

\section{DISCUSSION}
\subsection{The Universality of the IR LF \label{sect_dis1}}
The overall IR LFs of A3266 and the Coma cluster are similar both in shape and in normalization at $L_{IR}\ge10^{42.7}$ \ergss.
Since both of the observations cover about 70\% of the $R_{200}$ area of the clusters, field coverage is not a factor in the comparison \citep{Popesso06,Barkhouse07}.
\citet{Bai06} compared the IR LF of the Coma cluster with that of the local field and did not find a significant difference in \lstar.
The comparison of the faint end slope gives different results depending on which field IR LF is used, but generally no significant difference is found. 
Since the IR LF of A3266 is similar to the Coma IR LF, it confirms the constancy of the shape of the IR LF from clusters to the field at the bright end ($L_{IR}\ga10^{43}$ \ergss), even for very rich clusters with M$\sim 10^{15}~M_{\sun}$.

\citet{Cortese03,Cortese05} studied the UV LF of several local clusters and found that the shape of the LF of the star-forming galaxies in clusters is consistent with that of the field.
\citet{Iglesias02} constructed the $H\alpha$ LF of three local clusters and found it is consistent with the field $H\alpha$ LF at the bright end.
\citet{Christlein03} found that the R-band LF of star-forming galaxies in clusters is indistinguishable from that of the field.
\citet{DePropris03} also found that the B-band LF of the late-type galaxies in clusters is very similar to that of field galaxies.
These results suggest a universal shape of the LF of star-forming galaxies in very different environments.

A possible explanation for this universal shape is that the majority of the star-forming galaxies we see in the clusters are galaxies accreted from the field and their star formation has not been altered by the cluster environment yet. 
The accreted galaxies referred to here are not interloper galaxies whose redshifts happened to fall into the redshift range of the cluster but which are not physically bound to the cluster.
The normalization of the cluster LF is still 10 times larger than the field IR LF normalization multiplied by the physical scale corresponding to the cluster redshift range, indicating the contamination from such interlopers is small.
The fact that the velocities of the galaxies studied here fall within the maximum line-of-sight velocity requirement for a bound cluster member (see Fig.~\ref{f_phaseplot}) also confirms that they are not interlopers.

Although these galaxies are not interlopers and their projected radial distances are all within $R_{200}$, it might be that they are recently accreted and still lie outside of $R_{200}$, where some of the most important cluster environmental processes, e.g., ram-pressure stripping, tidal interaction, and starvation, have not started to affect them yet.
If we assume the mass function of the cluster galaxies does not change across the cluster and the cluster galaxy distribution follows the dark matter distribution, which can be modeled by a NFW profile \citep{Navarro97} with $R_{200}/r_{s}=6$ for a cluster at the mass range of A3266 \citep{Eke98}, we find that about 30\% of the galaxies projected within $R_{200}$ lie beyond $R_{200}$. 
However, many simulations \citep{Balogh00,Gill05} have shown that even for galaxies in the 1-2 $R_{200}$ radius, about 50\% are a 'backsplash' population, galaxies which have once penetrated into the $R_{200}$ region.
This implies that only about 15\% of the cluster members we studied here could really be recently accreted field galaxies that have never been inside of $R_{200}$.
Therefore, even if all of the recently accreted field galaxies are star-forming galaxies, they cannot entirely account for the 35\% of star-forming galaxies we detect in A3266 down to $M_{R}\le -19.15$ (23\% in Coma).

Given this evidence, we conclude that hot gas starvation is unlikely to alter the overall star formation properties of cluster galaxies.
This is because starvation starts to work around $R_{200}$ and it is a slow process ($\sim$Gyr).
If it does have a major effect on galaxy star formation, we would expect to see many star-forming galaxies that are in the transition stage of the process and they should have a lower SFR (and $L_{IR}^{*}$) compared to the field galaxies not affected by this process.
This contradicts the unchanged bright end IR LF shape we find from the field to the cluster. 
However, we need to note that given the accuracy of our measurement of the cluster IR LF, we are not sensitive to a variation of 30\% in $L^{*}_{IR}$.
In addition, the uncertainty of the field IR LFs further decrease the sensitivity of the comparison of environments.
So a modest change of $L^{*}_{IR}$ value between cluster and field galaxies might be buried by the uncertainties.

Novertheless, our result is compatible with certain classes of cluster environmental effects. 
For example, if an effect is limited to a smaller central region than $R_{200}$, it would be possible that these star forming galaxies are cluster members that have not penetrated deep inside the cluster, and therefore their star formation remains unaffected. 
For instance, ram-pressure stripping only works efficiently within 0.5 $R_{200}$ for a Milky Way-type galaxy \citep[e.g.,][]{Treu03}, typical in stellar mass of the galaxies of the star-forming galaxies in our study. 
Using the same NFW model, we can estimate that about 60\% of the galaxies with projected radial distance $<R_{200}$ lie beyond $0.5R_{200}$. 
Of the backsplash population in the 1-2 $R_{200}$ radius, about 90\% have penetrated into the $0.5R_{200}$ region \citep{Gill05}.
If we assume galaxies located between $0.5R_{200}$ and $R_{200}$ have the same fraction of the deep-penetrating backsplash population, we find about 30\% of galaxies with projected radial distance $<R_{200}$ have never been inside of the $0.5R_{200}$ region. 
This fraction is comparable to the star-forming galaxy fraction we find.
If we take into consideration that the deep-penetrating fraction in the 0.5-1 $R_{200}$ radius range may be higher than in the 1-2 $R_{200}$ radius range and the fact that not all field galaxies are star-forming galaxies to start with, then we still fall a little short to fully account for the star-forming galaxies detected in the cluster. 
However, because the effective ram-pressure stripping radius can be less than $0.5R_{200}$ for galaxies not in radial orbits \citep{Treu03}, the percentage of galaxies not affected by ram-pressure stripping can be larger than 30\% and therefore we can account for the majority of the star-forming galaxies we find in the cluster.

In addition, because the ram-pressure stripping works on a short time scale \citep[$\sim10^{7}$ yrs,][]{Abadi99} and can convert a star-forming galaxy into a quiescent one very quickly, there will be few galaxies in transition \citep{Balogh04,Verdugo08} and this will keep the shape of the IR LF unchanged while changing the fraction of star-forming galaxies. 
To keep the shape of IR LF unchanged, the ram-pressure stripping also needs to work equally effectively on the relevant range of galaxy luminosities \citep{Cortese03}.
This seems inconsistent with our understanding of ram-pressure stripping, which should be more efficient for the low-mass galaxies than for the high-mass ones.
But in the Coma cluster where we can constrain the faint end slope of the IR LF, we do see a flatter faint end slope in the core region, which may be due to the more efficient ram-pressure stripping for fainter galaxies.

As a conclusion, our results suggest that the cluster environment has little effect on the shape of the high-end cluster infrared luminosity function, but is characterized by a decrease of IR-bright galaxy fraction towards the cluster center. 
Nonetheless, the proportion of such galaxies is high enough to make it implausible that they are all recently accreted from the field. 
To be compatible with both the number of infrared-active galaxies and the behavior of the LF, any significant cluster environmental effect must be largely confined to the inner region (e.g., 0.5 $R_{200}$) and probably acts relatively quickly. 
These arguments would be compatible with effects through ram-pressure stripping or a similar process, but not with starvation. 

\subsection{The Universality of the Evolution of the IR LFs}
In addition to the universality of the shape of the bright-end IR LF from clusters to field at $z=0$, the evolution of this LF also shows universality from clusters to field. 
If this evolution is passive or driven by some mechanism that works in both environments, then it is no surprise that we see similar evolution in clusters and field. 
However, our observations are also compatible with some specific forms of differential evolution in the cluster environment that might arise from cluster-specific mechanisms such as ram pressure stripping. 
In general, slow mechanisms are unlikely because they would leave a transitional galaxy population in the process of evolving, and we would have to appeal to coincidence to explain the similarity in $L_{IR}^{*}$ between the field and the clusters.
However, as discussed in the preceding section, a fast-acting cluster environmental effect is possible, if the cluster keeps replenishing its star forming galaxies from the field.
In this case the cluster mechanism can quickly turn off the star formation in some of accreted galaxies, for example, those penetrate deep inside the cluster, leaving the $L_{IR}^{*}$ of rest of the star forming galaxies untouched. 
As a result, the observed $L_{IR}^{*}$ evolution in cluster will just follow the $L_{IR}^{*}$ evolution in the field, which, on the other hand, can be driven by some environmental mechanism not efficient in the cluster, e.g., galaxy merger.

The interpretation of the evolution of the IR LF normalization, however, is more complicated.
This normalization depends both on the accretion history and SF-suppression efficiency of the cluster. 
It also depends on the space density of the star-forming galaxies in the field, which is the reservoir for cluster accretion.
For example, assume rich clusters at high redshift have higher accretion rates than local clusters, as suggested by some hierarchical structure formation studies \citep[e.g.g][]{Kauffmann95}, and also have comparable SF-suppression efficiency to local clusters.
Then the evolution of the IR LF normalization would be stronger than that in the field, but would still more or less follow the field number density evolution. 
This behavior would be compatible with our results. 
However, given the large uncertainties in both the normalization evolution and the relevant parameters of high-z cluster behavior, this interpretation is very uncertain.

\section{CONCLUSIONS}
Despite its complicated dynamical state, the IR LF in A3266 is very similar to that of the Coma Cluster, suggesting a universal form of the bright end IR LF ($L_{IR}>10^{43}$ \ergss) in nearby rich clusters.
The shape of this cluster IR LF, at least at the bright end, is not significantly different from the field IR LF at the same redshift. 
The fraction of the IR-bright galaxies with $L_{IR}>10^{42.7}$ \ergss\ (SFR$> 0.2~M_{\sun}~yr^{-1}$) down to $M_{R}\le -20.15$ is $40^{+3}_{-4}\%$ in A3266 and $31^{+4}_{-5}\%$ in Coma.
There are few IR-bright galaxies within a projected radius of $r<0.3$ Mpc, and even they may not actually lie within the cores. 
The fraction of IR-bright galaxies in these two clusters increases linearly with radius, as $(0.24\pm0.02) + (0.36\pm0.06)R/R_{200}$.
The decrease of the blue galaxy fraction toward higher density regions only accounts for part of the trend, specifically, for the IR-bright galaxies with SFR$>1M_{\sun}~yr^{-1}$; the fraction of red galaxies with moderate SFR (0.2$~M_{\sun}~yr^{-1}<SFR<1~M_{\sun}~yr^{-1}$) also decreases with increasing galaxy density.
These results show that the cluster environment is characterized by a decrease of IR-bright galaxy fraction toward the cluster center, but not by a change in $L_{IR}^{*}$.
If some cluster environmental mechanism is responsible for the difference, this mechanism must be confined to the central regions (e.g., $< 0.5 R_{200}$) of the cluster and probably acts relatively quickly.
These properties would be compatible with a process like ram pressure stripping, but not with a slower-acting one such as starvation.

When comparing the A3266-Coma composite IR LF with that of two rich clusters at $z\sim0.8$, MS~1054 and RX~J0152, we find a strong evolution characterized by $L^{*}_{IR}\propto (1+z)^{3.2^{+0.7}_{-0.7}},\Phi^{*}_{IR}\propto (1+z)^{1.7^{+1.0}_{-1.0}}$.
This $L^{*}_{IR}$ evolution is indistinguishable from that in the field, and the $\Phi^{*}_{IR}$ evolution is stronger, but still consistent with that in the field.
The similarity of the evolution of bright-end IR LF in very different cluster and field environments suggests either this evolution is driven by the mechanism that works in both environments, or clusters continually replenish their star-forming galaxies from the field, yielding an evolution in the IR LF that is similar to the field.

The mass-normalized integrated SFR ($\Sigma SFR/M$) within 0.5$R_{200}$ ($\sim1.5$ Mpc) of A3266 and Coma are similar and show an evolution of $(1+z)^{5.3\pm1.2}$ compared with those of the two high-$z$ clusters. 
The evolution derived from these four rich clusters lies under a more scattered evolution trend traced by a larger cluster sample.
The large scatter indicates that in addition to evolution, cluster star formation may correlate with cluster masses, IGM properties, and perhaps other parameters.
However, for the four rich clusters studied by MIPS, neither the IR LFs nor $\Sigma SFR/M$ within $0.5R_{200}$ show a dependence on cluster mass, suggesting the mass dependence, if any, is weak for clusters with $M>\sim 4\times10^{14}~M_{\sun}$.

\acknowledgments
We thank the referee for insightful comments that help to improve the paper. 
This research has made use of the NASA/IPAC Extragalactic Database (NED) which is operated by the Jet Propulsion Laboratory, California Institute of Technology, under contract with the National Aeronautics and Space Administration.
This work was supported by funding for \textit{Spitzer} GTO programs by NASA,
through the Jet Propulsion Laboratory subcontracts \#1255094 and \#1256318.
\bibliographystyle{apj}

\clearpage
\LongTables
\input{tab1.tex}

\end{document}

%% file: tab1.tex
\begin{deluxetable}{rrrcrcc}
\tablecolumns{7}
\tablewidth{0pc}
\tablecaption{Cluster galaxies with 24 \micron\ emission $> 0.25$ mJy. \label{tbl1}}
\tablehead{
\colhead{RA} & \colhead{Dec} &\colhead{z} &\colhead{ref}&\colhead{$f_{24}$}&\colhead{log$L_{IR}$}&\colhead{SFR}\\
\colhead{J2000}&\colhead{J2000}&&&\colhead{mJy}&ergs $s^{-1}$ &$M_{\sun}$ yr$^{-1}$}
\startdata
  67.76958& -61.45722&0.055&1&  1.56$\pm$  0.12&43.24&   0.8\\
  67.80042& -61.45417&0.060&1&  2.64$\pm$  0.14&43.44&   1.2\\
  67.81625& -61.44028&0.068&1&  0.78$\pm$  0.09&42.98&   0.4\\
  67.78708& -61.41972&0.064&1& 14.59$\pm$  1.13&44.10&   5.7\\
  67.67458& -61.45444&0.063&1&  0.50$\pm$  0.10&42.82&   0.3\\
  67.81917& -61.50222&0.053&1&  1.00$\pm$  0.11&43.09&   0.6\\
  67.82167& -61.53083&0.053&2&  1.16$\pm$  0.10&43.14&   0.6\\
  67.73667& -61.53389&0.066&1&  0.52$\pm$  0.09&42.83&   0.3\\
  67.59000& -61.48028&0.065&1&  2.60$\pm$  0.15&43.43&   1.2\\
  67.95292& -61.43806&0.063&1&  0.29$\pm$  0.06&42.62&   0.2\\
  67.72583& -61.35083&0.062&1&  0.29$\pm$  0.08&42.62&   0.2\\
  67.68000& -61.35222&0.062&1&  1.38$\pm$  0.11&43.20&   0.7\\
  67.95958& -61.51278&0.058&1&  0.55$\pm$  0.08&42.86&   0.3\\
  67.94875& -61.38000&0.049&1& 38.16$\pm$  4.88&44.46&  12.9\\
  67.59167& -61.53417&0.056&1&  4.14$\pm$  0.18&43.61&   1.8\\
  67.74917& -61.57750&0.067&1&  0.48$\pm$  0.08&42.80&   0.3\\
  67.62667& -61.33972&0.059&2&  4.29$\pm$  0.24&43.62&   1.9\\
  67.49542& -61.41583&0.053&1&  2.24$\pm$  0.13&43.38&   1.1\\
  68.02125& -61.53528&0.060&1&  0.77$\pm$  0.11&42.98&   0.4\\
  67.57583& -61.58611&0.061&1&  4.53$\pm$  0.27&43.64&   2.0\\
  67.67458& -61.28583&0.056&1&  0.42$\pm$  0.08&42.75&   0.3\\
  67.94750& -61.29861&0.070&1&  0.71$\pm$  0.09&42.95&   0.4\\
  68.10458& -61.37917&0.056&1&  1.38$\pm$  0.12&43.20&   0.7\\
  67.39500& -61.46778&0.061&1&  0.52$\pm$  0.08&42.84&   0.3\\
  67.98708& -61.30778&0.059&1&  0.37$\pm$  0.06&42.72&   0.2\\
  68.12458& -61.51250&0.055&1&  0.65$\pm$  0.08&42.92&   0.4\\
  67.47708& -61.33056&0.063&1&  0.83$\pm$  0.10&43.01&   0.5\\
  67.42875& -61.33944&0.065&1&  0.31$\pm$  0.07&42.64&   0.2\\
  67.58000& -61.63056&0.056&2&  0.58$\pm$  0.07&42.88&   0.3\\
  67.87750& -61.64611&0.064&1& 11.94$\pm$  0.90&44.01&   4.6\\
  67.71375& -61.65639&0.066&1&  4.45$\pm$  0.19&43.64&   1.9\\
  67.77375& -61.23778&0.056&3&  2.10$\pm$  0.14&43.36&   1.0\\
  68.15708& -61.33278&0.059&1&  0.65$\pm$  0.11&42.92&   0.4\\
  67.29642& -61.48814&0.064&1&  0.91$\pm$  0.09&43.05&   0.5\\
  67.83042& -61.68472&0.059&1&  1.14$\pm$  0.11&43.13&   0.6\\
  67.91000& -61.21667&0.052&1&  0.47$\pm$  0.09&42.80&   0.3\\
  67.32076& -61.33136&0.061&2&  3.52$\pm$  0.19&43.56&   1.7\\
  68.27792& -61.43722&0.059&1&  0.66$\pm$  0.09&42.92&   0.4\\
  68.24375& -61.36028&0.070&1&  0.39$\pm$  0.06&42.73&   0.2\\
  67.25958& -61.40500&0.053&2&  0.91$\pm$  0.09&43.05&   0.5\\
  67.43542& -61.24917&0.058&2&  3.64$\pm$  0.16&43.58&   1.7\\
  68.23083& -61.31944&0.054&1&  0.89$\pm$  0.11&43.04&   0.5\\
  67.71583& -61.19333&0.057&1&  0.41$\pm$  0.09&42.75&   0.3\\
  67.91000& -61.70222&0.054&1&  0.79$\pm$  0.09&42.99&   0.4\\
  68.14125& -61.64417&0.065&1&  2.50$\pm$  0.16&43.42&   1.2\\
  68.28583& -61.35417&0.060&1&  4.60$\pm$  0.25&43.65&   2.0\\
  67.65667& -61.71389&0.056&1&  2.04$\pm$  0.11&43.35&   1.0\\
  68.18750& -61.27444&0.068&1&  1.62$\pm$  0.11&43.26&   0.8\\
  68.14208& -61.23306&0.058&1&  7.90$\pm$  0.35&43.86&   3.3\\
  68.36167& -61.38167&0.057&1&  0.77$\pm$  0.08&42.98&   0.4\\
  68.10292& -61.20167&0.066&1&  0.33$\pm$  0.07&42.67&   0.2\\
  67.36250& -61.68000&0.066&1&  0.56$\pm$  0.12&42.87&   0.3\\
  68.39000& -61.40750&0.060&1&  2.00$\pm$  0.13&43.35&   1.0\\
  67.39375& -61.20833&0.066&1&  2.19$\pm$  0.15&43.37&   1.1\\
  67.32917& -61.23417&0.059&1&  0.43$\pm$  0.08&42.77&   0.3\\
  67.82375& -61.75667&0.054&1&  1.28$\pm$  0.12&43.17&   0.7\\
  67.55375& -61.16167&0.065&1&  4.00$\pm$  0.20&43.60&   1.8\\
  67.38750& -61.19000&0.057&2&  0.32$\pm$  0.08&42.65&   0.2\\
  67.56833& -61.14333&0.063&1&  0.76$\pm$  0.11&42.97&   0.4\\
  67.41500& -61.17611&0.067&1&  1.91$\pm$  0.15&43.33&   1.0\\
  68.41417& -61.35056&0.066&1&  5.24$\pm$  0.18&43.69&   2.2\\
  67.32333& -61.20222&0.057&2&  0.62$\pm$  0.09&42.91&   0.4\\
  68.30417& -61.24806&0.056&1&  4.58$\pm$  0.25&43.64&   2.0\\
  67.27167& -61.22278&0.054&1&  1.43$\pm$  0.07&43.21&   0.7\\
  67.27250& -61.21778&0.063&1&  8.45$\pm$  0.42&43.88&   3.4\\
  68.28167& -61.22722&0.055&1&  0.31$\pm$  0.06&42.64&   0.2\\
  67.99600& -61.13322&0.058&1&  0.84$\pm$  0.10&43.01&   0.5\\
  67.77917& -61.10333&0.066&1&  0.40$\pm$  0.07&42.74&   0.2\\
  68.44750& -61.58944&0.057&1&  6.13$\pm$  0.27&43.76&   2.6\\
  67.77625& -61.80556&0.057&1&  2.32$\pm$  0.15&43.39&   1.1\\
  68.39958& -61.26389&0.065&1&  1.45$\pm$  0.11&43.21&   0.7\\
  68.23458& -61.17361&0.068&1&  4.13$\pm$  0.22&43.61&   1.8\\
  67.11667& -61.27222&0.066&1&  1.41$\pm$  0.11&43.20&   0.7\\
  68.30333& -61.20028&0.068&1&  8.80$\pm$  0.35&43.90&   3.5\\
  67.55708& -61.10250&0.059&1&  0.70$\pm$  0.10&42.94&   0.4\\
  67.89667& -61.09083&0.062&2&  4.28$\pm$  0.30&43.62&   1.9\\
  67.26042& -61.17694&0.056&2&  3.17$\pm$  0.13&43.51&   1.5\\
  68.53292& -61.49389&0.058&1& 11.05$\pm$  0.73&43.97&   4.2\\
  67.04500& -61.30389&0.057&1& 14.89$\pm$  0.33&44.11&   5.8\\
  66.98000& -61.54333&0.057&1&  0.44$\pm$  0.09&42.77&   0.3\\
  68.27792& -61.13306&0.059&1&  0.79$\pm$  0.13&42.99&   0.4\\
  67.09208& -61.69417&0.063&2&  0.62$\pm$  0.11&42.91&   0.4\\
  68.58500& -61.58056&0.058&1&  1.45$\pm$  0.12&43.21&   0.7\\
  66.94000& -61.31556&0.057&1& 50.79$\pm$  7.96&44.56&  16.2\\
  68.56500& -61.26556&0.067&1&  5.77$\pm$  0.21&43.73&   2.4\\
  68.67042& -61.41944&0.058&1&  0.95$\pm$  0.10&43.07&   0.5\\
  66.97875& -61.23306&0.055&1&  4.82$\pm$  0.26&43.66&   2.1\\
  68.65792& -61.55306&0.059&2&  0.91$\pm$  0.09&43.04&   0.5\\
  68.60250& -61.64861&0.069&1&  0.76$\pm$  0.09&42.97&   0.4\\
  68.71708& -61.44722&0.051&1&  0.39$\pm$  0.09&42.73&   0.2\\
  68.73083& -61.40361&0.053&1&  0.29$\pm$  0.06&42.61&   0.2\\
  67.27667& -61.03361&0.065&2&  1.51$\pm$  0.11&43.23&   0.8\\
  67.13542& -61.06556&0.056&2&  0.47$\pm$  0.09&42.80&   0.3\\
  66.75083& -61.37556&0.057&1&  1.18$\pm$  0.11&43.14&   0.6\\
  66.84458& -61.23528&0.059&1&  0.51$\pm$  0.10&42.82&   0.3\\
  66.73292& -61.44333&0.057&1&  2.26$\pm$  0.14&43.38&   1.1\\
  68.79542& -61.39861&0.057&1&  0.85$\pm$  0.08&43.01&   0.5\\
  68.80083& -61.48722&0.054&1&  1.26$\pm$  0.10&43.16&   0.7\\
  66.95458& -61.13778&0.056&1&  1.95$\pm$  0.15&43.34&   1.0\\
  67.24625& -61.01750&0.057&2&  1.50$\pm$  0.12&43.23&   0.8\\
  68.50125& -61.82722&0.054&1&  7.46$\pm$  0.36&43.85&   3.2\\
  68.83346& -61.52411&0.056&1&  2.35$\pm$  0.18&43.40&   1.1\\
  68.85875& -61.55722&0.054&1&  0.70$\pm$  0.12&42.95&   0.4\\
  66.60250& -61.35194&0.057&1& 21.20$\pm$  0.69&44.22&   7.5\\
  66.60667& -61.28972&0.056&1&  0.40$\pm$  0.23&42.74&   0.2\\
  66.66458& -61.20472&0.056&1&  1.30$\pm$  0.16&43.18&   0.7\\
  68.68708& -61.92250&0.058&1&  8.98$\pm$  0.71&43.90&   3.6\\
  68.63458& -61.94778&0.057&1&  3.19$\pm$  0.22&43.52&   1.5\\
  67.83171& -61.40511&0.047&1&  4.51$\pm$  0.25&43.64&   2.0\\
\enddata
\tablecomments{Table \ref{tbl1} is published in its entirety in the electronic edition of the {\it Astrophysical Journal}. A portion is shown here for guidance
regarding its form and content.\\
References for redshifts: (1) \citet{Christlein03};
			(2) \citet{Quintana96};
			(3) \citet{Green90}.}
\end{deluxetable}